\documentstyle[12pt,amssymbols]{article}
\input math_macros.tex

\begin{document}
\begin{titlepage}
\begin{center}
November 25, 1997     \hfill    LBNL- 40369 \\

\vskip .5in

{\large \bf  On Quantum Theories of the Mind}
\footnote{This work was supported by the Director, Office of Energy 
Research, Office of High Energy and Nuclear Physics, Division of High 
Energy Physics of the U.S. Department of Energy under Contract 
DE-AC03-76SF00098.}
\vskip .50in
Henry P. Stapp\\
{\em Lawrence Berkeley National Laboratory\\
      University of California\\
    Berkeley, California 94720}
\end{center}

\vskip .5in

\begin{abstract}

Replies are given to arguments advanced in this journal that claim to show 
that it is to nonlinear classical mechanics rather than quantum mechanics 
that one must look for the physical underpinnings of consciousness.

\end{abstract}
\end{titlepage}

\newpage
\renewcommand{\thepage}{\arabic{page}}
\setcounter{page}{1}

In a  paper with  the same  title as  this one  Alwyn  Scott (1996)  has  given
reasons for rejecting the idea that quantum theory will play an  important role
in understanding the connection  between brains and  consciousness. He suggests
that it is to  nonlinear  classical mechanics,  not quantum  mechanics, that we
should look for the  physical underpinnings of   consciousness. I shall examine
here all of his arguments, and show why each  one fails.

Scott contrasts, first,  the linearity of quantum  theory with the nonlinearity
of  certain  classical  theories, and  notes the   complexities  induced by the
latter. Thus he asks: ``Is not liquid  water essentially different from gaseous
hydrogen  and  oxygen?'' Of  course it is!  And this  difference  is generated,
according to  quantum field theory, by  certain  nonlinearities in that theory,
namely the  nonlinearities in  the coupled  {\it field  equations}. These field
equations (or, more generally,  Heisenberg equations) are the direct analogs of
the coupled nonlinear equations of the corresponding classical theory, and they
bring into quantum  theory the analogs  of the classical  nonlinearities: these
nonlinearities  are  in no way  obstructed  by the  linearity of  the {\it wave
equation}. 

To understand this point it is helpful  to think of the equation of motion  for
a  classical  statistical  ensemble.  It is  linear: the  sum of two  classical
statistical  statistical  ensembles  evolves into  the sum of  the two  evolved
ensembles. This linearity  property is a trivial  consequence of the  fact that
the elements  of the  ensembles are  imaginary  copies of one  single  physical
system, in different contemplated  states, and hence they do not  interact with
one  another.  Thus in   classical   statistical  mechanics we  have   both the
(generally)   nonlinear  equations  for  coupled  {\it  fields}, and  also  the
(always) linear equation for a certain statistical quantity. 

Similarly, in quantum field theory we have both the (generally) nonlinear field
equations  for the  coupled {\it  fields},  and also  the (always)  linear wave
equation for a certain statistical quantity, the {\it wave function}. The  fact
that a group of several atoms can  behave very differently from how they  would
behave if  each one  were alone  is a  consequence of  the  nonlinearity of the
field equations: this nonlinearity is not blocked by the linearity of the  wave
equation. 

This blurring of the  important distinction between  the completely  compatible
linear and  nonlinear aspects of  quantum theory is  carried over  into Scott's
discussion of solitons.  The nonlinear field  equations make the  parts of this
configuration of fields hang together  indefinitely, and never  spread out like
a  wave,  as  could  be   verified  by  doing    experiments  that    probe its
`togetherness'  by  making  several  measurements  simultaneously  at  slightly
separated  points: the various  simultaneously  existing parts  of  the soliton
never move  far  apart. There is  no  conflict between  this   stability of the
soliton and the  linearity of the  quantum mechanical   wave equation. The wave
function for the  {\it center-of-mass  of the  soliton}  does eventually spread
out in exactly the way  that   {\it a statistical  ensemble} consisting of the
{\it centers  of the solitons}   in an  ensemble of freely  moving solitons (of
fixed finite extension) would do:  the spreading out of the {\it wave function}
of the center-of-mass of  a  soliton just gives the  diffusion analogous to the
spreading out of a   statistical ensemble of  superposed {\it centers of mass},
due to the   distribution in  this ensemble of  velocities of  these centers of
mass:  the extended object itself, the  soliton, does not spread out; its parts
are   held  together  by a  nonlinear  effect  that  can be   attributed to the
nonlinearity  of the field equations.

This obscuring by  Scott of the  important conceptual  distinctions between the
two very   different aspects  of the  soliton  associated  with the  linear and
nonlinear aspects of  quantum theory creates, I  think, a very false impression
of  some   significant   deficiency  of   quantum  theory  with  regard  to the
manifestation of the analogs in  quantum theory of nonlinear classical effects.
No such   deficiency  exists: the  atoms  of  hydrogen and  oxygen do  combine,
according to quantum theory, to form water.

Failure carefully to follow through this conceptual distinction is 
the root of the failures of all of Scott's arguments.

Scott emphasizes the smallness of the spreading of the wave function of
the center-of-mass of Steffi Graf's tennis ball. That situation involves
the motion of a large massive object, the tennis ball, relative to, say,
a baseline on a large tennis court. 

A pertinent analogous  situation in the brain  involves the motion of a calcium
ion from the exit of a microchannel of diameter 1 nanometer to a target trigger
site for the release of a vesicle of neuro-transmitter into the synaptic cleft.
The irreducible  Heisenberg uncertainty in the  velocity of the ion as it exits
the  microchannel  is about  $1.5$  m/sec, which  is  smaller than  its thermal
velocity by a  factor of about $4  \times 10^{-3}$. The  distance to the target
trigger site is about  $50$ nanometers. So the  spreading of the wave packet is
of the order of $0.2$ nanometers,  which is of the order of the size of the ion
itself, and of  the target  trigger site. Thus  the decision  as to whether the
vesicle  is  released or  not, in an  individual   instance, will  have a large
uncertainty due to  the Heisenberg  quantum uncertainty in  the position of the
calcium ion relative to the trigger  site: the ion may hit the trigger site and
release the  vesicle, or it may miss  the trigger site and  fail to release the
vesicle. These two possibilities, yes or no, for the release of this vesicle by
this ion continue to  exist, in a superposed state,  until a ``reduction of the
wave  packet''  occurs.  Thus, if  there is a  part of the  wave  function that
represents a situation in which a  certain particular {\it set of vesicles} are
released,  due  to the   relevant  calcium  ions  having been   captured at the
appropriate sites, then  there will be other nearby  parts of the wave function
of the  brain  in  which some  or all  of the   relevant  captures do  not take
place---because, for this  part of the wave  function, some of the calcium ions
miss their target---and hence the corresponding vesicles are not released. 

This means, more generally, in a situation that corresponds to a very large 
number N of synaptc firings, that until a reduction occurs,  all of the 
$2^N$ possible combinations of firings and no firings will be represented 
with comparable statistical weight in the wave function of the brain/body 
and its environment.  Different combinations of these firings and 
no firings can lead to very different macroscopic behaviours of the body 
that is being controlled by the this brain, via the {\it highly nonlinear} 
neurodynamics of the brain. Thus the collapse effectively chooses between
very different possible macroscopic bodily actions.

I do not suggest that the mechanism  just cited, involving the diffusion of the
calcium ions in  the nerve terminals  is the {\it only}  sources of significant
differences between the  macroscopic consequences  of the quantum and classical
descriptions  of  brain  dynamics, for many  other  possible  effects have been
identified by quantum physicists  interested in brain dynamics. But this effect
is directly  computable, whereas some  of the others  depend on complex factors
that are not yet  under theoretical  control, and hence  could be challenged as
questionable.  But this effect  pertaining to  calcium ions in  nerve terminals
gives very  directly a reason for the  the  inappropriateness of the example of
Steffi  Graf's  tennis ball:  the  relevant  scales are  enormously  different.
Because  of  this the  huge   difference in  scales,  the   consequences of the
Heisenberg   uncertainty  principle,  and the  subsequent  collapses  that they
entail, are irrelevant to the outcome  of the tennis match, but are critical to
the bodily outcome of a brain activity that depends on the action at synapses.

Scott now lists a  number of reasons  for believing that  quantum theory is not
important  in  brain  dynamics in a  way that  would  relate to  consciousness.
However, as I shall now  explain, none of these  arguments has any relevance to
the issue, which hinges on a putative connection between conscious thoughts and
quantum reduction events. 

The point is this. The  quantum reduction/collapse  events mentioned above are,
according to orthodox Copenhagen  quantum theory, closely tied to our conscious
experiences. I  believe that all  physicists who suggest  that consciousness is
basically a  quantum aspect of nature  hold that our  conscious experiences are
tied to quantum collapses. The motivation for this belief is not merely that it
was only by adopting this idea that the founders of quantum theory were able to
construct  a  rational  theory  that  encompassed  in a  unified and  logically
coherent way the  regularities of physical  phenomena in both the classical and
quantum domains. The second powerful  motivation is that this association seems
provide a  natural physical  basis for the  unitary character  of our conscious
experiences. The point is that quantum  theory demands that the collapse of the
wave function  represent in  Dirac's words  ``our more precise  knowledge after
measurement''. But the  representation of the  increase in knowledge associated
with  say,  some    perception,  would be    represented in  the  brain  as the
actualization, as  a whole unit, of a  complex brain state  that extends over a
large part of  the brain.  Collapse events of  some kind are  necessary to make
ontological  sense out of  orthodox-type  quantum theory, and  these events can
never be pointlike  events: they must have finite  extension. But once they are
in principle   non-pointlike, they  need not be  tiny, and can  quite naturally
extend over an entire physical system.  The natural and necessary occurrence in
quantum theory of these  extended holistic  macroscopic realities that enter as
inseparable  and   efficacious units  into the  quantum   dynamics---and which,
according to the physical theory  itself, are associated with sudden increments
in our   knowledge---seems to  put the  physical and   psychological aspects of
nature into a much  closer and more  natural  correspondence with each other in
quantum theory  than in classical  mechanics, in which  every large-scale thing
is,  without  any  loss,   completely   decomposable,  both   ontologically and
dynamically, into its tiny parts.

Scott's  first reason  for claiming  quantum theory  to be  unimportant to mind
pertains to the speading of wave packets in molecular dynamics. That effect was
just considered, and the  crucial spreading of the  calcium ion wave packets in
nerve  terminals was shown to  be large  compared to the ion  size, contrary to
Scott's estimate.

Scott then  considers a  subject he  has worked on:  polarons.  He says the the
effect of the  quantum  corrections is to  degrade the global  coherence of the
classical polaron. But  this ``degrading'' is not  just some fuzzying-up of the
situation: it is the very thing that is of interest and importance here. In the
case of a body/brain this  ``degrading'' is, more  precisely, the separation of
the wave  function into  branches  representing  various  classical describable
possibilities.     However,  only one  of  these   classical   possibilities is
experienced in the mind associated  with this body/brain. Quantum theory in its
present  form is  mute on  the  question  of which  of these   possibilities is
experienced: only a statistical rule is provided. But then what is it that {\it
undoes}  this  huge (in  our case)   degrading that  the  linear wave  equation
generates. It is not the  classical nonlinearities,  for the quantum analogs of
these nonlinearities are built into  the part of the quantum dynamics that {\it
creates} the  superposition of the classically  describable possibilities: they
are built into  the  Schroedinger equation. A  collapse, which  is the putative
physical  counterpart of the conscious  experience, is a  different effect that
does not enter  into the Schroedinger  equation. Nor does  it enter at all into
the classical approximation to quantum  theory. In that approximation there are
no Heisenberg  uncertainties  or  indeterminacies, and hence  no collapses, and
hence from the persective of the encompassing and more basic quantum theory, no
physical counterparts of our conscious experiences.

His next two points concern the difficulty of maintaining ``quantum coherence''
in a warm, wet  brain. The  brain is a complex  structure with  built-in energy
pumps. The  question of  whether or not  long-range quantum  coherence could be
maintained is  difficult to settle  theoretically. Some  explorations have been
made (Vitiello, 1995), but the matter is not yet settled. On the other hand, my
theory yields  important  quantum effects that  are not wiped  out by decohence
effects  and that  could lead to  the  evolution of a  dynamically  efficacious
consciousness in  coordination with  evolution of brains  without requiring any
long-range quantum coherence (Stapp, 1997a,b). 

Scott's  next item is  the theory for  the  propagation of an  action potential
along a nerve fiber. He  points out that this  propagation is well described by
the  classical  Hodgkin-Huxley  equation.  But even  among  neuroscientists who
accept classical  mechanics as an adequate  foundation for brain dynamics there
is a  recognition  that although  in some  situations  the  parallel processing
structure produces  reliable and essentially  deterministic behaviors of groups
of  neurons, in  spite of  the   essentially  stochastic  character  of the the
distribution  of individual  pulses on the  individual  neurons, in other cases
there are long-range  correlations in the timings  of pulses. One can expect in
cases where thermal and other classical fluctuation effectively cancel, in such
a way as to give reliable and deterministic behaviors, that the quantum effects
associated with collapses will probably have no major macroscopic consequences.
But in cases where long-range  correlations of pulse timings arise, the precise
details of  these timings must  be controlled  in part by  stochastic variables
even in a  completely  classical model  that  generallly  conforms to, say, the
Hodgkin-Huxley equation. In these more  delicate situations there is ample room
for the  large-scale effects  associated with  quantum  collapses of brain-wide
quantum states to play a decisive  dynamical role {\it within} the framework of
possibilities compatible  with the classical  Hodgkin-Huxley equations. Indeed,
the actualization of  global brain states would be  expected produce fine-tuned
global regularities that classical mechanics could not account for.

Scott's  final point  is about   Schroedinger's cat. He  says the  Schroedinger
equation cannot be  constructed  because the cat does not  conserve energy. But
the usual assumption in these studies of the quantum mind-brain is that quantum
theory is universally valid, in the sense that the Schroedinger equation is the
equation of  motion for  the entire  universe, in  absence of  collapse events.
Partial systems are  defined by integrating over  the other degrees of freedom,
and their energies are not conserved.

\newpage
{\bf References}

Fogelson, A.L. \& Zucker, R.S. (1985),`Presynaptic calcium diffusion
from various arrays of single channels: Implications for transmitter
release and synaptic facilitation', {\it Biophys. J.}, {\bf 48}, 
pp. 1003-1017.

Scott, A. (1996), `On quantum theories of the mind', {\it Journal of
Consciousness Studies}, {\bf 6}, No. 5-6, pp.484-91.

Stapp, H. (1993), {\it Mind, Matter, and Quantum Mechanics},
(Berlin: Springer), Chapter 6.

Stapp, H. (1997a), `Pragmatic Approach to Consciousness'
  To be published in {\it The Neural Correlates of Consciousness},
  ed., N. Osaka; To be re-published in {\it Brain and Values}, ed.
  K. Pribram, Lawrence Erlbaum, Mahwah, NJ.
\newline Availaible at http://www-physics.lbl.gov/$\sim$stapp/stappfiles.html

Stapp, H. (1997b) `Quantum ontology and mind-matter synthesis',
  in {\it The X-th Max Born Symposium}, eds., P. Blanchard and A. Jadczyk,
  to be published by Springer-Verlag, Berlin.
\newline Availaible at http://www-physics.lbl.gov/$\sim$stapp/stappfiles.html

Vitiello, G (1995), `Dissipation and memory capacity in the quantum brain
model', Int. J. Mod. Phys. {\bf B9}, 973-89.

Zucker, R.S. \& Fogelson, A.L. (1986), `Relationship between transmitter 
release and presynaptic calcium influx when calcium enters through disrete
channels', {\it Proc. Nat. Acad. Sci. USA}, {\bf 83}, pp. 3032-3036.

\end{document}